\documentclass[twocolumn,superscriptaddress,aps,preprintnumbers,amsmath,amssymb,prd,nofootinbib]{revtex4-1}

\usepackage{graphicx}
\usepackage{dcolumn}
\usepackage{hyperref}
\usepackage{color}
\usepackage{bm}
\usepackage{siunitx}
\usepackage{braket}
\usepackage{amsfonts,amsmath,amssymb,bm,tensor}
\usepackage{comment}
\usepackage{ifpdf}
\usepackage{slashed}
\usepackage[mathscr]{eucal}
\usepackage[utf8]{inputenc}
\usepackage{cancel}
\usepackage{enumerate}
\usepackage{tikz}
\usepackage{tikz-3dplot}
\usepackage{subfigure}
\usepackage{physics}
\usepackage{multirow}
\usepackage{fourier}

\definecolor{dgreen}{rgb}{0.05, 0.70, 0.05}
\definecolor{rossoferrari}{HTML}{D9073D}
\definecolor{mediumblue}{HTML}{0000CD}
\definecolor{forestgreen}{HTML}{228B22}
\definecolor{desy_blue}{HTML}{009EE2}
\definecolor{desy_orange}{HTML}{FD8800}
\definecolor{light_pink}{rgb}{1,0.4,0.4}
\definecolor{light_blue}{rgb}{0.284602,0.317763,0.963947}
\hypersetup{setpagesize=false,bookmarksnumbered=true,bookmarksopen=true,%
colorlinks=true,linkcolor=light_blue,urlcolor=rossoferrari,citecolor=rossoferrari,linktocpage=false}

\begin{document}


\title{
SU(\texorpdfstring{$N$}{})-natural inflation
}
\author{Tomohiro Fujita}
\affiliation{Waseda Institute for Advanced Study, Waseda University, Shinjuku, Tokyo 169-8050, Japan}
\affiliation{Research Center for the Early Universe, The University of Tokyo, Bunkyo, Tokyo 113-0033, Japan}
\author{Kyohei Mukaida}
\affiliation{Theory Center, IPNS, KEK, 1-1 Oho, Tsukuba, Ibaraki 305-0801, Japan}
\affiliation{Graduate University for Advanced Studies (Sokendai), 1-1 Oho, Tsukuba, Ibaraki 305-0801, Japan}
\author{Kai Murai}
\affiliation{ICRR, University of Tokyo, Kashiwa, 277-8582, Japan}
\affiliation{Kavli IPMU (WPI), UTIAS, University of Tokyo, Kashiwa, 277-8583, Japan}
\author{Hiromasa Nakatsuka}
\affiliation{ICRR, University of Tokyo, Kashiwa, 277-8582, Japan}

\begin{abstract}
	We study the SU($N$) gauge fields coupled with the inflaton through the Chern--Simons coupling and propose a general procedure for constructing homogeneous, isotropic, and attractor solutions of the gauge fields during inflation.
	Gauge fields develop various VEVs corresponding to different spontaneous symmetry breaking patterns of SU($N$), where embedded SU($2$) subgroups are broken with the spatial rotation SO($3$) symmetry.
	As specific examples, we construct the stable solutions for $N = 3$ and $4$.
	We numerically solve the gauge field dynamics and confirm that our analytic solutions are complete and attractor.
	Notably, the proposed approach is applicable to the other simple Lie groups.
\end{abstract}

\preprint{KEK-TH-2349}

\maketitle

\paragraph*{\textbf{Introduction.---}}%
Cosmological inflation successfully describes the primordial Universe by addressing the horizon/flatness problems and providing the seeds of anisotropies observed in the cosmic microwave background (CMB).
The hot Universe is expected to have been generated after inflation; this requires coupling between the inflaton and other sectors.
However, the flatness of the inflaton potential should be protected from radiative corrections to ensure a sufficient duration of inflation.
Axion-like particle is an attractive candidate for the inflaton~\cite{Freese:1990rb,Adams:1992bn,Kim:2004rp} because its (approximate) shift symmetry controls flatness.
Moreover, its derivative couplings can induce rich phenomena, including the amplification of gauge fields~\cite{Ratra:1991bn,Garretson:1992vt}.

The model wherein the inflaton is coupled with SU$(2)$ gauge fields through the Chern--Simons coupling, termed as chromo-natural inflation~\cite{Adshead:2012kp}, has attracted significant attention.
In the presence of an inflaton velocity, the gauge fields have a \textit{homogeneous and isotropic attractor} solution~\cite{Maleknejad:2011jw,Maleknejad:2013npa,Domcke:2018rvv,Wolfson:2020fqz,Wolfson:2021fya} that is suitable for explaining the current Universe.
Moreover, the backreaction from the gauge fields enables slow-roll inflation with a sub-Planckian axion decay constant.
Such background gauge fields induce linear couplings between the metric and gauge field tensor perturbations, and significant chiral gravitational waves are produced~\cite{Cook:2011hg,Dimastrogiovanni:2012ew}.
Although the original chromo-natural inflation is excluded from CMB observations owing to the overproduction of gravitational waves~\cite{Adshead:2013qp,Adshead:2013nka},
this conflict can be circumvented by introducing additional fields~\cite{Obata:2014loa,Obata:2016tmo,Maleknejad:2016qjz,Dimastrogiovanni:2016fuu,Adshead:2016omu,DallAgata:2018ybl}.

The SU$(2)$ gauge group is the simplest example of gauge fields with a homogeneous and isotropic attractor solution. In general, we can extend the chromo-natural inflation model to an SU($N$) gauge group with $N>2$.
We refer to this extension as ``SU($N$)-natural inflation.'' 
Previous discussions on the SU($N$) case were often limited to the simplest choice of the SU(2) subgroup; it was considered that the SU($N$) case simply leads to chromo-natural inflation~\cite{Adshead:2012kp,Maleknejad:2012fw}.
Alternatively, certain studies have considered the $N/2$ copies of the SU(2) subgroups in SU($N$) and discussed the production of gravitational waves~\cite{Caldwell:2016sut,Caldwell:2017chz,Caldwell:2018feo}.
However, as shown in this paper, most homogeneous and isotropic attractor solutions for the SU($N$)-natural inflation remain unexplored.

Herein, we propose a general procedure for determining homogeneous and isotropic attractor solutions for the SU($N$)-natural inflation under three reasonable assumptions.
We identify new solutions whose gauge field amplitudes exceed those in chromo-natural inflation.
Furthermore, the effective potential implies that
the solution with the largest amplitude is the most stable solution for SU$(N)$ with $N>2$.
Each solution corresponds to a non-equivalent embedding of the SU$(2)$ subgroup in SU$(N)$; this implies different spontaneous symmetry breaking patterns of SU$(N)$ under the chemical potential of the Chern--Simons number from a non-vanishing inflaton velocity.
We also perform numerical simulations of the gauge field dynamics and confirm that the analytically determined solutions are attractor and complete for $N = 3$ and $4$.

\smallskip
\paragraph*{\textbf{Model.---}}%
We consider a pseudo-scalar inflaton $\phi$ coupled with SU($N$) gauge fields $A_\mu^a$ through Chern–-Simons coupling:
\begin{align}
    \mathcal{L}=
        -\frac{1}{4}F_{\mu\nu}^a F^{a \mu\nu}
        +\frac{1}{2}\partial_\mu \phi \partial^\mu \phi
        -V(\phi)
        +\frac{\phi}{4f}F_{\mu\nu}^a \tilde{F}^{a \mu\nu}.
\end{align}
The field strength of the SU($N$) gauge fields $F^a_{\mu\nu}$ and its dual $\tilde{F}^{a\mu\nu}$ are defined as
\begin{align}
    F^a_{\mu\nu}
    \equiv
    \partial_\mu A^a_\nu - \partial_\nu A^a_\mu - g f^{a b c} A^b_\mu A^c_\nu,
    \quad
    \tilde{F}^{a \mu \nu}
    \equiv
    \frac{\epsilon^{\mu \nu \rho \sigma} F^a_{\rho \sigma}}{2\sqrt{-\tilde{g}}},
\end{align}
where $f^{a b c}$ denotes the structure constant of the SU($N$) algebra, and the superscripts $a,b,$ and $c$ range from $1$ to $N^2-1$.
$g$ denotes the gauge coupling constant, and $\tilde{g}$ represents the determinant of the spacetime metric.
We assume that the background metric during inflation is well approximated by the de Sitter universe,
$\mathrm{d}s^2 = \mathrm{d} t^2 - a(t)^2  \mathrm{d}\bm{x}^2$, with an exponentially growing scale factor,
$a(t)\propto e^{H t}$, and a constant Hubble parameter, $H=\mathrm{const}$.

\smallskip
\paragraph*{\textbf{Ansatz and solutions.---}}%
We focus on non-trivial attractor solutions of the SU($N$) gauge fields under a non-zero inflaton velocity during inflation.
To identify such solutions, we adopt three assumptions, that is, the generalized electric and magnetic fields in SU($N$) are (i) homogeneous,
(ii) static, and
(iii) parallel.

First, we consider homogeneous solutions of the gauge fields, $A_i^a(t,\bm{x}) = A_i^a(t)$, because the observed Universe is highly homogeneous.
We adopt the temporal gauge, $A_t^a = 0$, and redefine the gauge field into a dimensionless one:
\begin{equation}
    M_i^a(t) \equiv \frac{g}{a(t) H}A^a_i(t)
    \qquad (i=x,y,z).
    \label{eq: A stable}
\end{equation}
The equation of motion (EoM) for the gauge fields can be expressed as
\begin{align}
    &\frac{\ddot{M}_i^a}{H^2} + \frac{3}{H}\dot{M}_i^a + 2M_i^a
    \nonumber\\
    &
    + f^{b a c} f^{b d e} M_j^c M_i^d M_j^e
    - \xi \epsilon_{i j k} f^{a b c} M_j^b M_k^c = 0,
    \label{eq: eom dot m}
\end{align}
where $\xi \equiv \dot{\phi}/(2 f H)$ characterizes the energy transfer from the rolling inflaton to the gauge fields.
Instead of solving the inflaton dynamics with the backreaction, we assume that $\xi$ remains constant for simplicity.

Second, we assume that the gauge fields are static, with the aim of determining stable attractor solutions.
The energy density of homogeneous gauge fields is expressed as
\begin{equation}
    \rho_A
    =
    \frac{H^4}{4g^2}
    \left[
        f^{a b c} f^{a d e} M_i^b M_j^c M_i^d M_j^e
        +2 \left( \frac{\dot{M_i^a}}{H} + M_i^a \right)^2
    \right].
\end{equation}
This becomes constant in time if $\dot{M}_i^a = 0$.
From this perspective, we assume $\dot{M}_i^a = 0$ as the staticity of the gauge fields.
Subsequently, the EoM \eqref{eq: eom dot m} is reduced to
\begin{equation}
    2M_i^a
    + f^{b a c} f^{b d e} M_j^c M_i^d M_j^e
    - \xi \epsilon_{i j k} f^{a b c} M_j^b M_k^c
    =
    0.
    \label{eq: static EoM of M}
\end{equation}
Here, we introduce the generalized electric and magnetic fields as
\begin{align}
    E^a_i
    &\equiv
    -F_{0 i}^a
    =
    -\frac{a H^2}{g}M_i^a,
    \label{eq: electric field}
    \\
    B_i^a
    &\equiv
    \frac{1}{2}\epsilon_{i j k} F^a_{j k}
    =
    -\frac{a^2 H^2}{2g} \epsilon_{i j k} f^{a b c} M_j^b M_k^c.
    \label{eq: magnetic field}
\end{align}
Note that, under homogeneity, $B_i^a$ only consists of the non-linear terms in the field strength.

Finally, we assume that the electric and magnetic fields are parallel in the gauge internal space.
Without energy injection, the gauge fields would be quickly diluted by cosmic expansion,
and the amplitude would decay as $M_i^a \propto a^{-1}$.
To sustain their amplitudes, coupling with an energy source, (\textit{i.e.}, inflaton), is crucial; this is the Chern--Simons coupling in our case.
This coupling can be rewritten as
\begin{equation}
    \sqrt{-\tilde g}\,\frac{\phi}{4f}F_{\mu\nu}^a \tilde{F}^{a \mu\nu}
    =
    -\frac{\phi}{f} \left(E_x^a B_x^a + E_y^a B_y^a + E_z^a B_z^a\right).
\end{equation}
Here, we regard $E_i^a$ and $B_i^a$ for $i = x, y, z$ as vectors with $N^2-1$ components. Thereafter, the inflaton is coupled with their inner products.
We expect that only their parallel components are sustained in the inflationary Universe and that the non-parallel components, if any, disappear quickly.
Thus, we assume that the electric and magnetic fields are parallel along each spatial direction.
To elaborate on this assumption, we restore the SU($N$) generators $T^a$ satisfying $\Tr (T^a T^b) = \delta^{a b}/2$ and also define $E_i\equiv E_i^a T^a$ and $B_i\equiv B_i^a T^a$.
We can rephrase the third assumption as $E_i \propto B_i \propto \mathcal{T}_i$ for $i = x, y, z$. Here, $\mathcal{T}_i \equiv n_i^a T^a$ is a linear combination of $T^a$ with $\Tr (\mathcal{T}_i^2) = 1/2$.

Using the abovementioned three assumptions, we derive the condition for gauge field configuration.
Based on the first and second assumptions, the electromagnetic fields satisfy
\begin{equation}
    B_i = ig \epsilon_{ijk} [E_j, E_k]/H^2.
\end{equation}
Combining this with the third assumption, $E_i \propto B_i \propto \mathcal{T}_i$, we find that $\mathcal{T}_i$ satisfies an SU(2) subalgebra
\begin{align}
    [\mathcal{T}_i, \mathcal{T}_j]
    =
    i \lambda \epsilon_{ijk} \mathcal{T}_k,
    \qquad
    \Tr \left( \mathcal{T}_i \mathcal{T}_j \right)
    =
    \frac{\delta_{i j}}{2},
    \label{eq: normalization of mathcalT}
\end{align}
up to a proportionality constant $\lambda$.
Without loss of generality, we can consider $\lambda$ as positive
by choosing the sign of $n_i^a$.
These equations can be recast as follows:
\begin{align}
   n_i^a n_j^b f^{abc}
   =
   \lambda \epsilon_{ijk}n_k^c,
   \qquad
   n_i^a n_j^a
   =
   \delta_{i j}.
   \label{eq: n orthogonality}
\end{align}
$\lambda$ represents the difference in the normalization between generators.
The original generator of SU($N$), $T^a$ or $\mathcal{T}_i$, is normalized as $\Tr (\mathcal{T}_i \mathcal{T}_j) = \delta_{ij}/2$.
However, Eq.~\eqref{eq: normalization of mathcalT} implies that not $\mathcal{T}_i$ itself but $\mathcal{T}_i/\lambda$ has the structure constant of the SU(2) generator $\epsilon_{ijk}$.
$\lambda$ reflects this difference, and its value depends on the choice of the SU(2) subalgebra in SU($N$).

Consequently, we obtain the gauge field configuration:
\begin{align}
    M_i^a = \sigma_i n_i^a
    \quad
    (i = x,y,z),
    \label{eq: Msigman}
\end{align}
where $n_i^a$ satisfies Eq.~\eqref{eq: n orthogonality}, and $\sigma_i$ represents the gauge field amplitude along each spatial direction.
Using this configuration, we can further simplify the EoM \eqref{eq: static EoM of M} as follows:
\begin{equation}
    2\sigma_i
    + \lambda^2 \sigma_i \sum_{l \neq i} \sigma_l^2
    - \xi \lambda \sum_{j,k} |\epsilon_{ijk}| \sigma_j \sigma_k
    =
    0
    \quad
    (i = x,y,z).
\end{equation}
This EoM leads to $\sigma \equiv \sigma_x = \sigma_y = \sigma_z$, and it has three solutions:
\begin{equation}
    \sigma
    =
    0, \ \frac{\sigma_-}{\lambda}, \ \frac{\sigma_+}{\lambda},
    \quad
    \mathrm{with}
    \quad
    \sigma_{\pm} \equiv \frac{\xi \pm \sqrt{\xi^2 - 4 }}{2}.
    \label{eq: sigma formula}
\end{equation}
Here, we omit the solutions involving opposite signs for two $\sigma_i$ as $\sigma_x = -\sigma_y = -\sigma_z = \sigma$, because this difference corresponds to a gauge transformation of $M_i^a$.

\smallskip
\paragraph*{\textbf{Properties of the solutions.---}}%
Next, we confirm the isotropy of the obtained solutions.
The electromagnetic fields are isotropic when they are rotationally invariant up to the gauge transformation.
As $E_i^a \propto M_i^a$, this condition on the electric field is equivalent to the isotropy of $M_i^a$, which is formulated as follows:
For an arbitrary rotation $R$, there exists a gauge transformation $G$ satisfying
\begin{align}
    R_{i j} M_j^a = G^{a b} M^b_i,
    \label{eq: isotropy condition}
\end{align}
where $R_{i j}$ is a rotation matrix in the three-dimensional space, and $G^{a b}$ is an SU($N$) gauge transformation acting on $M_i^a$.
The isotropy of the magnetic field is also ensured by Eq.~\eqref{eq: isotropy condition}, and thus, it is necessary and sufficient for the electromagnetic fields to be isotropic.
Notably, the obtained solutions~\eqref{eq: Msigman} with $\sigma_i = \sigma$ are isotropic.
As $\mathcal{T}_i$ generates an SU(2) subalgebra, we can always determine the gauge transformation $G$ generated by the SU(2) subalgebra, which cancels an arbitrary spatial rotation $R$ owing to the isomorphism between the SU(2) and SO(3) algebras.

The background amplitude (or VEV) of the gauge fields spontaneously breaks the spatial rotation symmetry and the global gauge symmetry of the SU($2$) subgroup into the diagonal SO($3$), namely SO($3$)$\times$ SU($2$) $\to$ SO($3$), similar to chromo-natural inflation.
On identifying an SU($2$) subgroup embedded in the SU($N$) group and determining its $n_i^a$ and $\lambda$, we can obtain the gauge field solution.
However, the expressions of the generator $\mathcal{T}_i$ and its component $n_i^a$ are altered by a gauge transformation.
Henceforth, we concentrate on $\lambda$, which is a gauge-independent quantity.
The solutions for static gauge fields with different values of $\lambda$ have genuinely different configurations.
For instance, the difference in $\lambda$ can be observed in the energy density of the gauge fields, which can be expressed as
\begin{align}
    \rho_A
    = \frac{3H^4}{2g^2} \xi \lambda \sigma^3
    \propto \lambda^{-2}.
\end{align}
Note that the slow-roll solution exists only when the energy density of the gauge fields is subdominant compared to that of $\phi$.
Thus, the duration of inflation can be different depending on the value of $\lambda$.
However, the difference of $\lambda$ can be compensated by a choice of other parameters such as $g$.
Furthermore, as described later, a different choice of SU$(2)$ subgroups characterized by $\lambda$ leads to a different breaking pattern of SU($N$).

We can derive the general expression for $\lambda$ in the case where the $\mathbf{m}$ representation of SU(2) is embedded in SU($N$).
When the $m \times m$ part of the fundamental representation $\mathbf{N}$ of SU($N$) corresponds to the $\mathbf{m}$ representation of SU(2) with $m = 2, 3, \ldots, N$,
the generators of the SU(2) subalgebra can be expressed with one spin operator, $T_z^{(\bm{3})}$, and two ladder operators, $T_x^{(\bm{3})} \pm i T_y^{(\bm{3})}$.
Particularly, the spin operator, which is equal to either one of $\mathcal{T}_i/\lambda$, can be represented by an $N \times N$ diagonal matrix:
\begin{equation}
    \mathrm{diag}
    \left[
        \frac{m-1}{2}, \frac{m-3}{2}, \ldots, -\frac{m-1}{2}, 0, \ldots, 0
    \right].
    \label{eq: m spin operator}
\end{equation}
By normalizing this matrix as an SU($N$) generator, we obtain
\begin{equation}
    \lambda = \left[ \frac{m (m^2-1)}{6} \right]^{-1/2}.
    \label{eq: lambda formula}
\end{equation}

\begin{table}[tbp]
    \centering
    \caption{Decomposition of fundamental and adjoint representations of SU(3) and SU(4)~\cite{ramond_2010}.
    }
    \label{tab: decomposition}
    \begin{tabular}{c c c c}
        $N$ & Subgroup & $\mathbf{N}$ & $\mathbf{N^2-1}$ 
        \\
        \hline
        3& SU(2)$\times$U(1) & $\bm{2}_{-1} + \bm{1}_2$ & $\bm{3}_0 + \bm{2}_3 + \bm{2}_{-3} + \bm{1}_0$ 
        \\
        & SU(2)& $\bm{3}$ & $\bm{3} + \bm{5}$ 
        \\
        4 & SU(3)$\times$U(1) & $\bm{3}_{-1} + \bm{1}_3$ & $\bm{8}_0 + \bm{3}_{-4} + \bar{\bm{3}}_4 + \bm{1}_0$ 
        \\
        & SU(2) & $\bm{4}$ & $\bm{3} + \bm{5} + \bm{7}$ 
        \\
        & 
        SU(2)$\times$SU(2) & $(\bm{2}, \bm{1}) + (\bm{1},\bm{2})$ & $(\bm{3}, \bm{1}) + (\bm{1}, \bm{3})+ (\bm{2}, \bm{2}) + (\bm{2}, \bm{2}) + (\bm{1}, \bm{1})$
        \\
        &  & $(\bm{2}, \bm{2})$ & $(\bm{3}, \bm{3}) + (\bm{3}, \bm{1})+ (\bm{1}, \bm{3})$
    \end{tabular}
\end{table}

\smallskip
\paragraph*{\textbf{Examples.---}}%
As concrete examples, we enumerate the solutions for the SU(2), SU(3), and SU(4)
gauge fields by following the proposed method explained thus far.
Moreover, if multiple independent SU(2) subgroups are embedded in the SU($N$),
the sum of the solutions of these SU(2) subgroups
satisfies the EoM.
Thereafter, the amplitude of the total solution is given by the root sum square of the individual amplitudes, because the individual solutions are orthogonal.

\smallskip
\paragraph*{\bf{SU(2).---}}%
First, we investigate the simplest case, $N=2$.
Clearly, the only SU(2) subgroup is SU(2) itself, which has $\lambda = 1$.
Therefore, we obtain
$    \sigma
    =
    0, \,\sigma_{\pm}.
$

\smallskip
\paragraph*{\bf{SU(3).---}}%
Next, we investigate the case of $N=3$.
We can consider two types of embeddings for SU(2) into SU(3), as presented in Table.~\ref{tab: decomposition}; these correspond to $m = 2$ and $3$ in Eq.~\eqref{eq: lambda formula}, respectively.
Therefore, we obtain
$    \sigma
    =
    0, \,
    \sigma_{\pm}, \,
    2 \sigma_{\pm}.
    \label{eq: SU(3) solutions}
$

\smallskip
\paragraph*{\bf{SU(4).---}}%
Finally, we investigate the case of $N=4$.
The embedding of the subgroups in SU(4) has four patterns, as presented in Table~\ref{tab: decomposition}.
The first pattern, SU(3)$\times$U(1) $\subset$ SU(4), has the same solutions as those in 
the case of SU(3).

The second pattern, SU(2) $\subset$ SU(4), corresponds to $m = 4$ in Eq.~\eqref{eq: lambda formula} and yields
$    \sigma = \sqrt{10} \sigma_{\pm}.$

The third embedding, SU(2)$\times$SU(2) $\subset$ SU(4), where $\bm{15} = (\bm{3},\bm{1})+(\bm{1},\bm{3})+(\bm{2},\bm{2})+(\bm{2},\bm{2})+(\bm{1},\bm{1})$,
contains two independent SU(2)'s.
Both correspond to $m = 2$ in Eq.~\eqref{eq: lambda formula}.
Including their superposition, we obtain
$    \sigma = 0, \sigma_{\pm}, \sqrt{2}\sigma_{\pm}, \sqrt{(\sigma_+)^2+(\sigma_-)^2}.$

The last pattern is SU(2)$\times$SU(2) $\subset$ SU(4), where $\bm{15} = (\bm{3},\bm{1})+(\bm{1},\bm{3})+(\bm{3},\bm{3})$.
The two SU(2) subgroups are spanned by
$\mathcal T_i^{(\bm{3},\bm{1})} \equiv \sigma_i \otimes 1_{2\times2}/(2\sqrt{2})$
and $\mathcal  T_i^{(\bm{1},\bm{3})} \equiv 1_{2\times2} \otimes \sigma_i/(2\sqrt{2})$.
As both of these have $\lambda = 1/\sqrt{2}$,
including their superposition, we obtain
$    \sigma = 0, \sqrt{2}\sigma_{\pm}, 2\sigma_{\pm}, \sqrt{2 \qty[(\sigma_+)^2+(\sigma_-)^2 ] }.
    \label{eq: sigma of SU(4) (ii) (3,1) and (1,3)}
$

\begin{figure}[t]
  \includegraphics[clip,width=0.9\columnwidth]{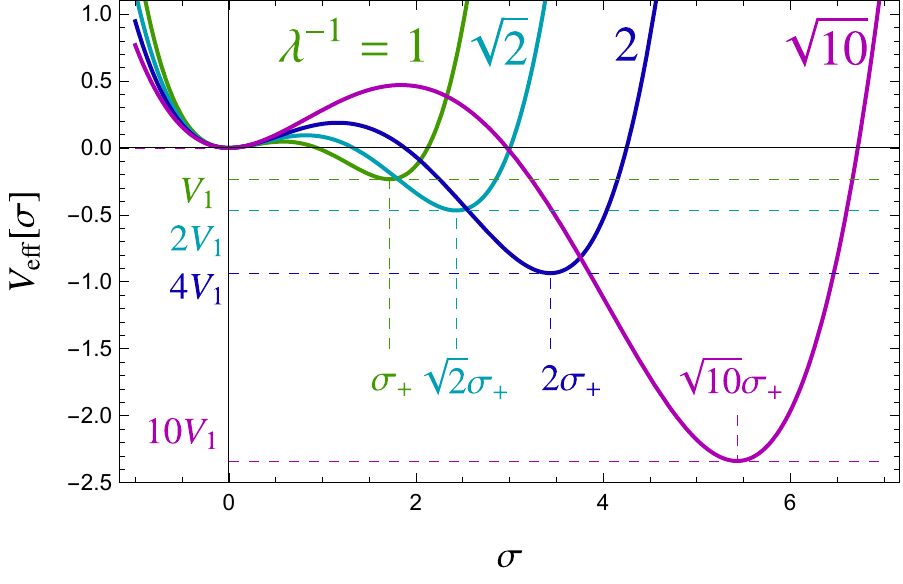}
  \caption{Effective potential $V_\mathrm{eff}(\sigma)$ for $\xi=2.3$ with $\lambda^{-1}=1$ (purple), $\sqrt{2}$ (violet),
  $2$ (orange), and $\sqrt{10}$ (green), which are solutions for the case of SU(3) or SU(4). The origin $\sigma=0$ is meta-stable, $\sigma_-/\lambda$ is the local maximum, and
  $\sigma_+/\lambda$ is the minimum with depth proportional to $\lambda^{-2}$.
 }
  \label{fig: Veffective}
\end{figure}

\smallskip
\paragraph*{\textbf{Stability.---}}%
Thus far, all the solutions satisfying our assumptions have been considered.
Here, we discuss the stability of solutions by introducing the effective potential of $\sigma$,
\begin{equation}
V_\mathrm{eff}(\sigma) = \frac{1}{2}\sigma^2-\frac{1}{3}\xi\lambda \sigma^3 +\frac{1}{4}\lambda^2 \sigma^4,
\end{equation}
whose extrema correspond to the solutions~\eqref{eq: sigma formula}.
The solutions at $\sigma=\sigma_-/\lambda$ are unstable because $V_\mathrm{eff}''(\sigma_-/\lambda)$ is negative for $\xi>2$.
By contrast, $V_\mathrm{eff}''(0)$ is always positive, and $V_\mathrm{eff}''(\sigma_+/\lambda)$ is also positive for $\xi>2$.
A comparison of the depth of the potential reveals that the trivial solution is the global minimum $V_\mathrm{eff}(0)<V_\mathrm{eff}(\sigma_+/\lambda)$ for $\xi<3/\sqrt{2}$.
However, for a larger energy injection $\xi>3/\sqrt{2}\approx 2.12$, the solution at the origin becomes meta-stable
and the non-trivial stable solution becomes the true vacuum, $V_\mathrm{eff}(0)>V_\mathrm{eff}(\sigma_+/\lambda)$.
Fig.~\ref{fig: Veffective} depicts the effective potential for $\xi=2.3$.
A non-trivial stable solution with a larger amplitude (smaller $\lambda$) has a deeper potential and is more energetically favored.
Thus, the original chromo-natural solution with $\lambda=1$ is not the most stable attractor for $N>2$.
Consequently, by eliminating the unstable solutions with $\sigma_-/\lambda$, we obtain stable solutions for SU(3) and SU(4) with
the amplitudes
\begin{align}
    &\mathrm{SU}(3):\quad \sigma = 0,\ \sigma_+,\ 2\sigma_+,
    \label{eq: SU(3)sols}\\
    &\mathrm{SU}(4):\quad \sigma = 0,\ \sigma_+,\ \sqrt{2}\sigma_+,\ 2\sigma_+,\ \sqrt{10}\sigma_+.
    \label{eq: SU(4)sols}
\end{align}

Next, we investigate the local stability of the solutions~\eqref{eq: SU(3)sols} and \eqref{eq: SU(4)sols} against general homogeneous perturbations~%
\footnote{%
    The stability of the solutions with $m = N$ against homogeneous perturbations has been investigated using a different approach~\cite{Adshead:2013nka}.
    However, they do not discuss the other solutions or their different amplitudes.
}.
The local stability of a solution is described by the Hessian matrix of the effective potential, which is obtained by differentiating the left-hand side of the EoM~\eqref{eq: static EoM of M} with respect to $M^b_j$:
\begin{align}
    S_{i j}^{a b}
    =
    &2 \delta_{i j} \delta^{a b}
    + \left(f^{e a b} f^{e c d} - f^{e a d} f^{e b c} \right) M_i^c M_j^d
    \nonumber \\
    &+ f^{f a c} f^{f b d} M_k^c M_k^d \delta_{i j}
    - 2\xi \epsilon_{i j k}f^{a b c} M_c^k.
\end{align}
As each set of $(i,a)$ and $(j,b)$ represents the component of the field space,
the Hessian of the effective potential $S_{i j}^{a b}$ is recognized as the $3(N^2-1) \times 3(N^2-1)$ matrix.
If the eigenvalues of $S_{i j}^{a b}$ are all non-negative for a certain solution of $M$, the solution is locally stable.
We verified that all the solutions, ~\eqref{eq: SU(3)sols} and \eqref{eq: SU(4)sols}, are locally stable.
Note that some eigenvalues naturally vanish corresponding to the gauge transformation of the solution.

\smallskip
\paragraph*{\textbf{Numerical results.---}}%
To confirm that our analytic stable solutions are dynamical attractors,
we numerically solve the EoM~\eqref{eq: eom dot m} of the homogeneous gauge fields for SU(3) and SU(4).
In the numerical calculations, although we retain the simplifying assumptions of the de Sitter spacetime $H=\mathrm{const}.$, constant rolling inflaton $\xi=\mathrm{const}.$, and homogeneity of the gauge fields, we do not impose the other two assumptions, staticity and parallelism.

The initial condition at $t=0$ is fixed as $\dot A_i^a(0) = 0$ to ensure that the gauge constraints are satisfied.
Using the Gaussian distribution, we let each of the $3(N^2-1)$ components $M_i^a(0)$ have a random initial amplitude with the variance $\Sigma^2$, which varies within $0<\Sigma^2\le 100/(N^2-1)$.
To depict the results, we introduce the dynamical mean amplitude of the gauge fields as
\begin{equation}
   \tilde \sigma(t)^2
    \equiv
    \frac{1}{3}\sum_{a,i} \qty[M_i^a(t)]^2.
    \label{eq: mean amplitude}
\end{equation}
We expect that $\tilde \sigma(t)$ converges to stable solutions in Eqs.~\eqref{eq: SU(3)sols} and \eqref{eq: SU(4)sols}.

\begin{figure}[htpb]
  \includegraphics[clip,width=0.9\columnwidth]{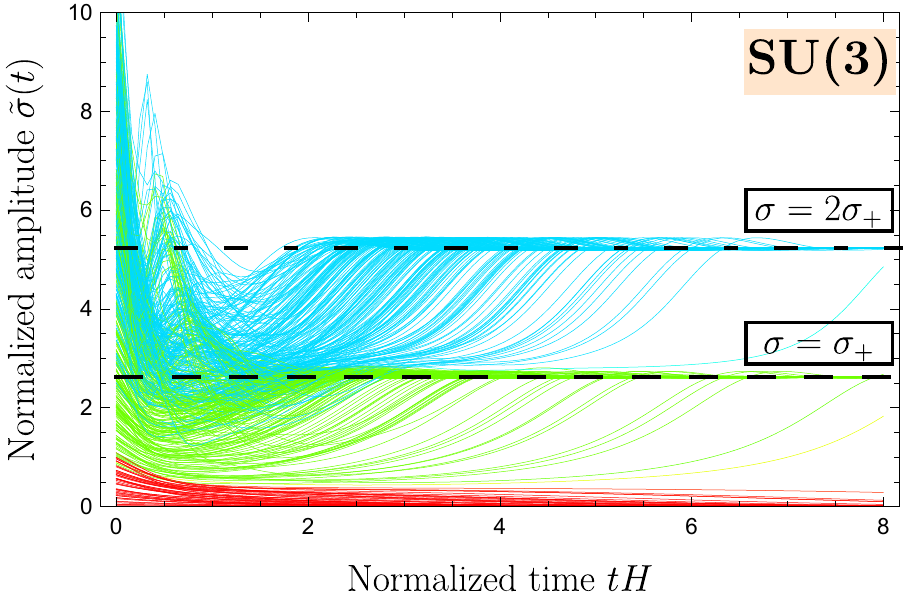}
  \includegraphics[clip,width=0.9\columnwidth]{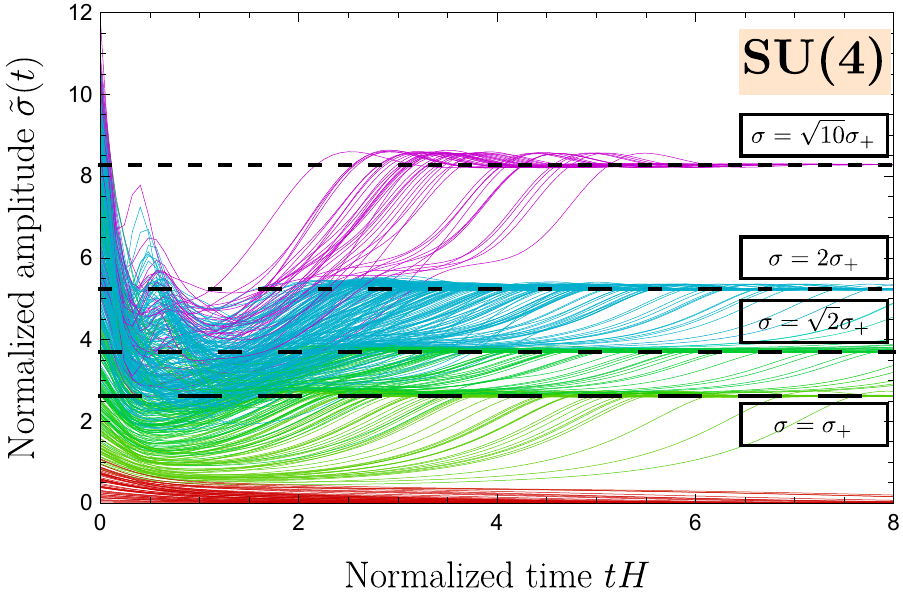}
  \caption{
  Time evolution of gauge field mean amplitude $\tilde \sigma(t)$ with $\xi = 3$ in the SU(3) (top) and SU(4) (bottom) cases.
  The initial conditions range over $\tilde \sigma(0) \simeq [0,10]$ with random amplitudes and orientations of the gauge fields.
  The black horizontal lines represent the analytic stable solutions in Eqs.~\eqref{eq: SU(3)sols} and \eqref{eq: SU(4)sols}, which act as attractors.
   }
  \label{fig: evol sigma}
\end{figure}

In Fig.~\ref{fig: evol sigma}, we depict 500 realizations of the gauge field dynamics for $\xi=3$. Each line represents the time evolution of $\tilde \sigma(t)$, and the colors denote the field values at the end of the calculation, $Ht=8$.
Three stable values exist in the SU($3$) case (the top panel), whereas five stable values exist in the SU($4$) case (the bottom panel), including the trivial solution $\sigma=0$.
As expected, $\tilde \sigma(t)$ is attracted to the stable solutions
and converges typically within a few Hubble times.
A realization with a larger initial amplitude tends to settle at the larger stable solution.
Upward transitions between the stable solutions can also be observed, which can be explained by the depth of the effective potential in Fig.~\ref{fig: Veffective}.

\smallskip
\paragraph*{\textbf{Discussion.---}}%
In this work, we extend the chromo-natural inflation model with the SU($N$) gauge group, termed as SU($N$)-natural inflation, and provide a general procedure for constructing isotropic attractor solutions of the gauge fields under three assumptions: the homogeneity, staticity, and parallelism of electromagnetic fields. The gauge field amplitudes of the constructed solutions are generally equal to or larger than those in chromo-natural inflation, and the maximum amplitude increases as $N^{3/2}$ for large values of $N$.
As specific examples, we construct the stable solutions for $N = 3$ and $4$.
We numerically solve the gauge field dynamics with numerous initial conditions and determine that all the constructed stable solutions are realized, while no other stable solutions are observed.
Our procedure can be straightforwardly applied to SU($N$) with $N \geq 5$ or other simple Lie groups.
Therefore, we can consider $A$- to $G$-natural inflation.

Several interesting extensions of our work can be explored.
We demonstrate the convergence of the gauge fields to stable solutions in Fig.~\ref{fig: evol sigma}; however, the relation between the initial condition and final configuration remains unknown.
Moreover, the time evolution of the gauge fields can be investigated by relaxing the assumptions of $H, \xi=\mathrm{const}.$
For example, the variation in $\xi$ may further induce a transition between the stable solutions for gauge fields.
To explore such gauge field behaviors, dedicated numerical simulations are required to solve the coupled dynamics of the inflaton and gauge fields.

Based on our background solutions, the linear perturbations of gauge fields can also be studied. In chromo-natural inflation, the power spectrum of gravitational waves sourced by the gauge field perturbation exponentially depends on the amplitude of the homogeneous gauge fields~\cite{Adshead:2013qp}.
Therefore, the larger amplitudes of the gauge fields in SU($N$)-natural inflation may significantly affect the properties of the sourced gravitational waves.
Lastly, we expect that some properties of the perturbations reflect the symmetry breaking pattern of the homogeneous solution.
For example, in the SU(3) case, the solutions with m$=2$ still possess the U(1) symmetry, whereas the solution with m$=3$ has no global gauge symmetry other than the diagonal SO(3) broken from SO(3)$\times$ SU(2).
It would be fascinating to explore how this difference in the symmetry breaking patterns reflects in the perturbations and eventually appears in observational signatures.

\medskip
\begin{acknowledgments}
We would like to thank Katsuki Aoki, Antonio De Felice, Masahiro Ibe, Kohei Kamada, Masahiro Kawasaki, and Yota Shamoto for the useful discussions and comments.
KaM was supported by the World Premier International Research Center Initiative (WPI Initiative), MEXT, Japan, and the Program of Excellence in Photon Science.
KyM was supported by the MEXT Leading Initiative for Excellent Young Researchers Grant Number JPMXS0320200430.
HN was supported by the Advanced Leading Graduate Course for Photon Science.
This work was supported in part by the Japan Society for the Promotion of Science (JSPS) KAKENHI, Grant Number JP18K13537 (TF), JP20J20248 (KaM), JP19J21974 (HN).
\end{acknowledgments}

\small
\bibliographystyle{apsrev4-1}
\bibliography{Ref}

\appendix


\section{Numerical calculation of the gauge fields}
\label{Subsup: numerical calculation}

Here, we describe some details pertaining to the numerical calculations.
We numerically solve the equations of motion (EoMs) for all components of the gauge fields, $A^a_i(t)$.
We use the Gell--Mann matrices as the basis of the adjoint representation of the SU(3) Lie algebra, and the matrices $\lambda_a \equiv 2T_a$ in Table.~\ref{fig_su4_basis} as that of SU(4).

\begin{table*}[htpb]
\caption{
  Basis of SU(4) gauge fields.
}
\label{fig_su4_basis}
\begin{align}
    &\lambda_1 =
    \begin{pmatrix}
    0&1&0&0\\
    1&0&0&0\\
    0&0&0&0\\
    0&0&0&0
    \end{pmatrix}
    ,
    &&\lambda_2 =
    \begin{pmatrix}
    0&-i&0&0\\
    i&0&0&0\\
    0&0&0&0\\
    0&0&0&0
    \end{pmatrix}
    ,
    &&&\lambda_3 =
    \begin{pmatrix}
    1&0&0&0\\
    0&-1&0&0\\
    0&0&0&0\\
    0&0&0&0
    \end{pmatrix}~
    &&&&\lambda_4 =
    \begin{pmatrix}
    0&0&1&0\\
    0&0&0&0\\
    1&0&0&0\\
    0&0&0&0
    \end{pmatrix},
    &&&&&\lambda_5 =
    \begin{pmatrix}
    0&0&-i&0\\
    0&0&0&0\\
    i&0&0&0\\
    0&0&0&0
    \end{pmatrix},
    \nonumber
    \\
    &\lambda_6 =
    \begin{pmatrix}
    0&0&0&0\\
    0&0&1&0\\
    0&1&0&0\\
    0&0&0&0
    \end{pmatrix},
    &&\lambda_7 =
    \begin{pmatrix}
    0&0&0&0\\
    0&0&-i&0\\
    0&i&0&0\\
    0&0&0&0
    \end{pmatrix},
    &&&\lambda_8 =
    \frac{1}{\sqrt{3}}
    \begin{pmatrix}
    1&0&0&0\\
    0&1&0&0\\
    0&0&-2&0\\
    0&0&0&0
    \end{pmatrix},
    &&&&\lambda_9 =
    \begin{pmatrix}
    0&0&0&1\\
    0&0&0&0\\
    0&0&0&0\\
    1&0&0&0
    \end{pmatrix},
    &&&&&\lambda_{10} =
    \begin{pmatrix}
    0&0&0&-i\\
    0&0&0&0\\
    0&0&0&0\\
    i&0&0&0
    \end{pmatrix},
    \nonumber
    \\
    &\lambda_{11} =
    \begin{pmatrix}
    0&0&0&0\\
    0&0&0&1\\
    0&0&0&0\\
    0&1&0&0
    \end{pmatrix},
    &&\lambda_{12} =
    \begin{pmatrix}
    0&0&0&0\\
    0&0&0&-i\\
    0&0&0&0\\
    0&i&0&0
    \end{pmatrix},
    &&&\lambda_{13} =
    \begin{pmatrix}
    0&0&0&0\\
    0&0&0&0\\
    0&0&0&1\\
    0&0&1&0
    \end{pmatrix},
    &&&&\lambda_{14} =
    \begin{pmatrix}
    0&0&0&0\\
    0&0&0&0\\
    0&0&0&-i\\
    0&0&i&0
    \end{pmatrix},
    &&&&&\lambda_{15} =
    \frac{1}{\sqrt{6}}
    \begin{pmatrix}
    1&0&0&0\\
    0&1&0&0\\
    0&0&1&0\\
    0&0&0&-3
    \end{pmatrix}.
    \nonumber
\end{align}
\end{table*}

SU($N$) gauge fields have $3(N^2-1)$ degrees of freedom, considering the three spatial directions and the $N^2-1$ components in the internal space.
The time components of the gauge fields, $A^a_t$, are non-dynamical variables; this results in $N^2-1$ gauge constraints, as follows:
\begin{align}
    0=\frac{\delta\mathcal L}{\delta A^a_t}
    =
    -\sum_{b,c}
    g f^{abc} \dot A^b_iA^c_i .
\end{align}
Thus, the gauge conditions are trivially satisfied when the initial condition is considered as $\dot A^a_i =0$ for all $a$.

\begin{figure}[htpb]
  \includegraphics[clip,width=0.9\columnwidth]{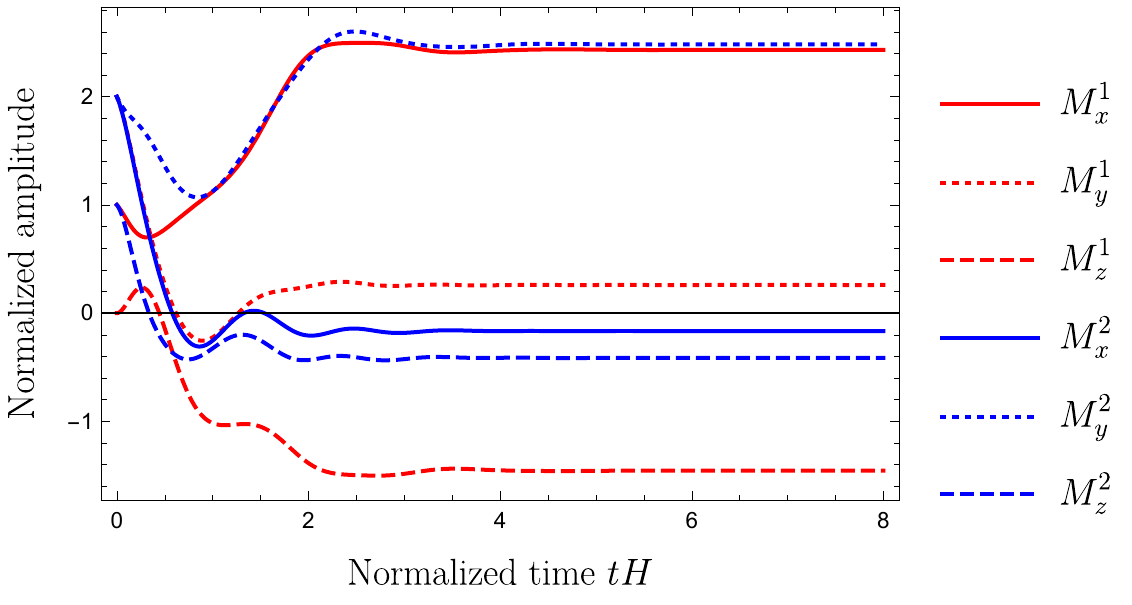}
  \caption{
  Typical time evolution of gauge field amplitudes with $\xi = 3$ in the SU(3) case.
  Only the components for $a=1, 2$, and $i=x,~y,~z$ are shown as examples.
  }
  \label{fig_field_value}
\end{figure}

The typical dynamics of the normalized field values $M_i^a(t) \equiv \frac{g}{a(t) H}A^a_i$ in the SU(3) case are depicted in Fig.~\ref{fig_field_value}.
Here, we adopt the following initial condition for the 24 components, as an example:
\begin{align}
    M_x^a &= (1,2,1,3,0,0,4,0),
    \\
    M_y^a &= (2,2,1,0,3,0,0,4),
    \\
    M_z^a &= (0,1,2,0,0,3,0,0).
\end{align}
During the first few Hubble times, the field values exhibit wiggling behaviors, following which they converge to stable values.

\begin{figure}[htpb]
  \includegraphics[clip,width=0.9\columnwidth]{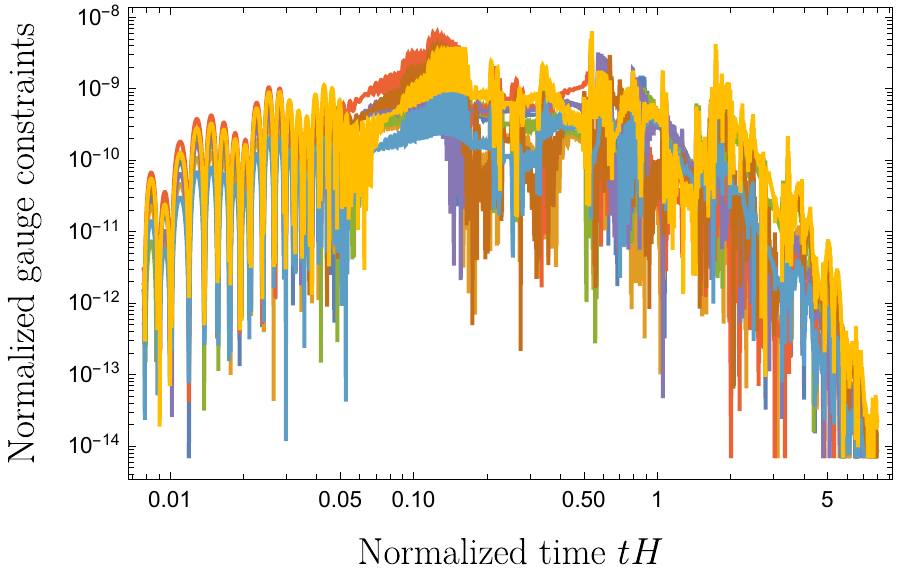}
  \caption{
  Errors in gauge constraints, Eq.~\eqref{eq_normalized_gaugeconst}, under the same dynamics as in Fig.~\ref{fig_field_value}.
  Eight lines corresponding to the gauge constraints from $A^a_0$ are plotted.
  The violation of gauge constraints is always smaller than $10^{-8}$.
  }
  \label{fig_gauge_const}
\end{figure}

EoMs conserve the gauge constraints; we numerically assess whether the gauge constraints are satisfied with high accuracy, as shown in Fig.~\ref{fig_gauge_const}.
The vertical axis represents the normalized gauge constraints, defined by
\begin{align}
    \frac{
    |\sum_{b,c} f^{abc} \dot A^b_iA^c_i|
    }{
    \sum_{b,c}|f^{abc} \dot A^b_iA^c_i|
    }.
    \label{eq_normalized_gaugeconst}
\end{align}

\begin{figure}[htpb]
    \includegraphics[clip,width=0.9\columnwidth]{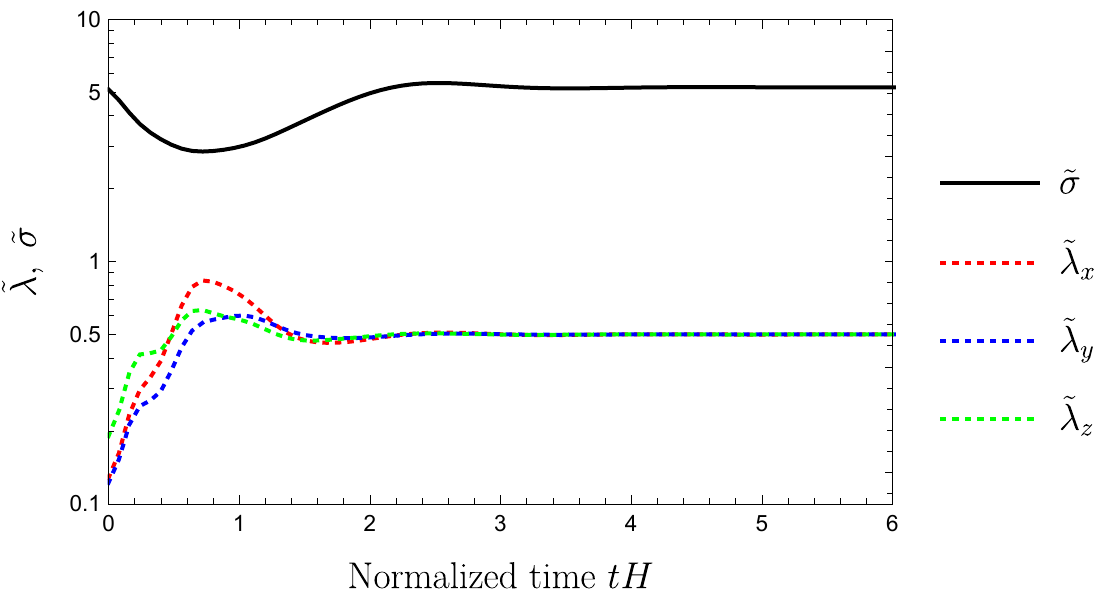}
  \includegraphics[clip,width=0.9\columnwidth]{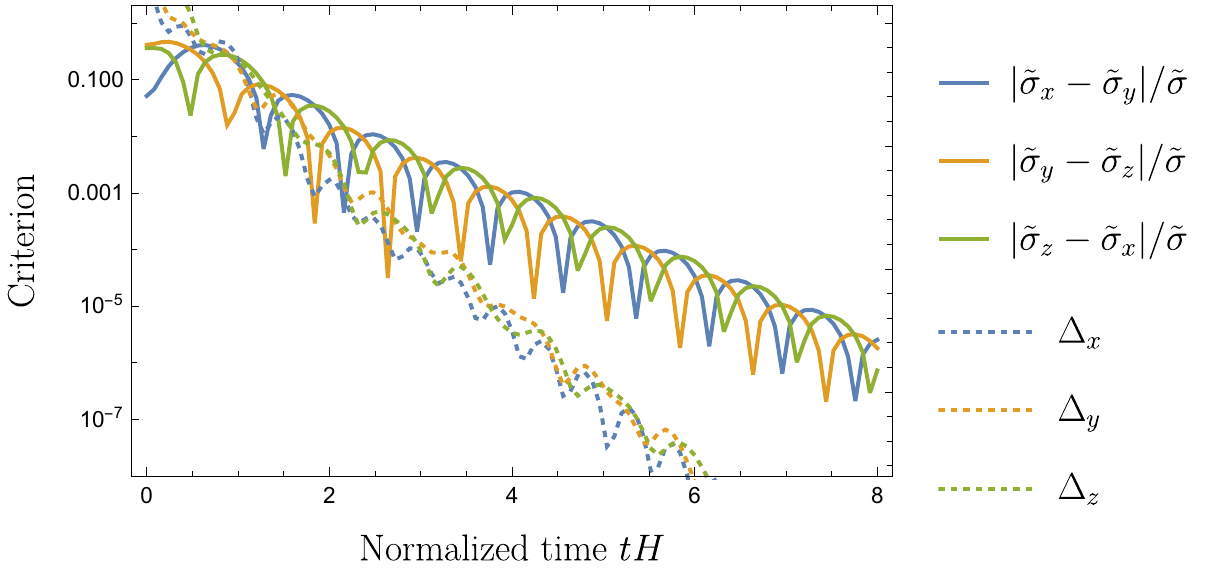}
  \caption{
  Convergence of gauge fields with $\xi = 3$ in SU(3).
  The top panel depicts the convergence of the amplitude and the coefficient of the commutation relation in Eqs.~\eqref{eq:crit iso} and~\eqref{eq:crit su2}.
  We also present the criterion on convergence for isotropy and the SU(2) subgroup in the bottom panel.
  }
  \label{fig_convergence}
\end{figure}

Furthermore, we investigate whether the numerical solutions satisfy the expected properties, that is, the isotropy and SU(2) subgroup, where the latter ensures that the electric and magnetic fields are parallel.
Here, we use the matrix form of the gauge fields, $M_i \equiv M_i^a T^a \propto A_i^a T^a$.
First, we discuss the isotropy of the gauge field configuration.
The amplitude of the $i-$component is expressed as
\begin{align}
    \tilde \sigma_i(t)
    &\equiv
    \sqrt{ 2 \mathrm{Tr}\left[ (M_i)^2 \right] }\,
    \label{eq:crit iso}
\end{align}
where the summation over $i$ is not considered.
Their root mean square reproduces the mean amplitude~\eqref{eq: mean amplitude} as $\tilde \sigma (t)=\sqrt{(\tilde \sigma_x^2 +\tilde \sigma_y^2 +\tilde \sigma_z^2)/3}$.
In the isotropic configuration, the three amplitudes $\tilde \sigma_i$ are equal, and $|\tilde \sigma_i-\tilde \sigma_j|/\tilde \sigma$ disappears.
The bottom panel in Fig.~\ref{fig_convergence} shows that $|\tilde \sigma_i-\tilde \sigma_j|/\tilde \sigma$ decays rapidly, and thus, the configuration becomes isotropic.

\begin{figure}[htpb]
     \includegraphics[clip,width=0.8\columnwidth]{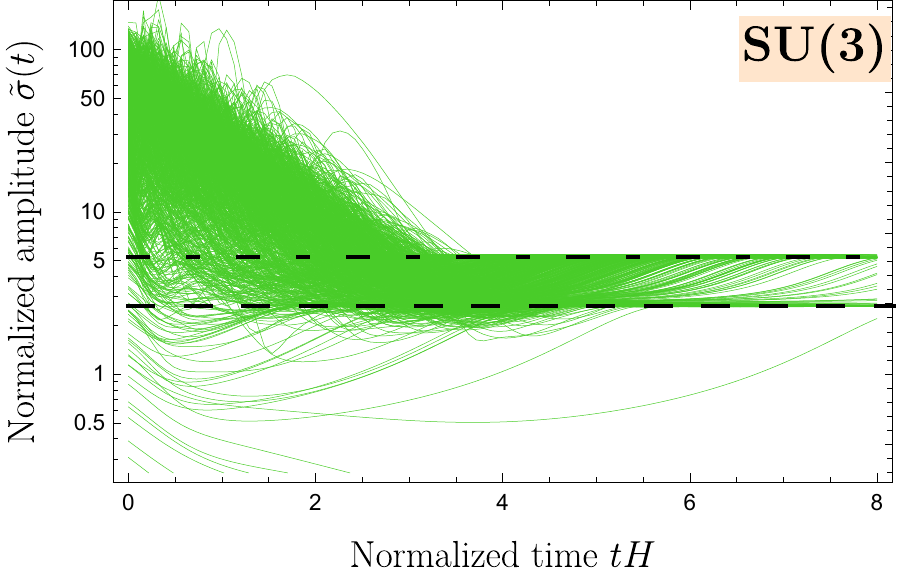}
  \includegraphics[clip,width=0.8\columnwidth]{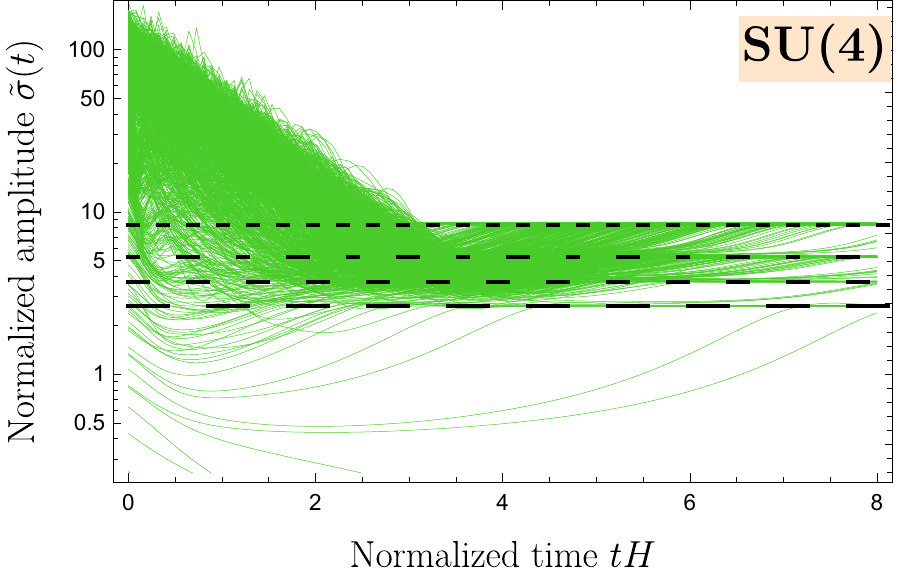}
  \caption{
  Dynamics of the mean-field amplitude with the extended initial conditions for $\xi=3$.
  Each green line represents the time evolution of the gauge fields under different initial conditions.
  The initial mean amplitudes range from $\tilde \sigma(t_0) \simeq [0,10^2]$.
  The black horizontal lines represent the stable solutions derived in this work.
  No additional attractor solution is identified.
  }
  \label{fig_initial_condition}
\end{figure}

Next, we assess whether the gauge fields lie in an SU(2) subgroup, which is characterized by the following quantities:
\begin{align}
    \Delta_x
    &=
    \frac{
    1
    }{\Tr((M_x)^2)}
    \Tr\left[ \left(
    M_x -
    \frac{[M_y,M_z]}{i\tilde \sigma_x \tilde \lambda_x}
    \right)^2 \right]
    ,
    \nonumber
    \\
    \tilde \lambda_x(t)
    &\equiv
    \frac{2}{i}
    \frac{\Tr(M_x[M_y,M_z]  )}{(\tilde \sigma_x)^3}
    .
    \label{eq:crit su2}
\end{align}
Furthermore, we define $\Delta_y$ and $\Delta_z$ using the permutation of $\{x,y,z\}$ in $\Delta_x$.
When $M_i$ reaches the stable solution, $\tilde \lambda_i(t)$ converges to $\lambda $ introduced in Eq.~\eqref{eq: normalization of mathcalT}.
When the magnetic field $B_x\propto [M_y, M_z]$ is proportional to the electric field $E_x\propto M_x$, $\Delta_x$ disappears.
The top panel in Fig.~\ref{fig_convergence} illustrates the time evolution of the mean amplitude $\tilde \sigma$ (solid line) and $\tilde \lambda_i$ (dotted lines) for $\xi=3$.
$\tilde\sigma$ and $\lambda_i$ converge to $2\sigma_+\simeq 5.2$ and $1/2$, respectively; this is consistent with our analytic estimation for the $m=3$ solution.
Moreover, $\Delta_i$ decays rapidly, as depicted in the bottom panel of Fig.~\ref{fig_convergence}.
Thus, we verified that the gauge fields converge to the isotropic SU(2) configuration.

To ensure that no more attractor solutions exist,
we extend the maximum value of the initial amplitude by a factor of 10
and implement 1000 realizations for both the SU(3) and SU(4) cases.
In contrast to the range of the initial condition $\tilde\sigma(0)\simeq [0,10]$ in Fig.~\ref{fig: evol sigma}, the initial amplitude $\tilde \sigma$ varies
within approximately $[0,100]$, as shown in Fig.~\ref{fig_initial_condition}.
Even with these thorough searches, we found no more stable solutions other than those derived analytically.

\end{document}